\documentclass[journal]{IEEEtran}

\ifCLASSINFOpdf
 
\else
  
\fi

\usepackage{graphicx}
\usepackage{geometry} 
\usepackage[export]{adjustbox}
\usepackage{float}
\usepackage{amsmath}
\usepackage{caption}
\usepackage{xcolor}
\usepackage{supertabular,booktabs}
\usepackage{setspace}
\usepackage{pdflscape} 
\usepackage{listings}    

\usepackage{graphicx}
\usepackage{multirow}% http://ctan.org/pkg/multirow
\usepackage{hhline}
\usepackage{array}
\newcolumntype{L}{>{\centering\arraybackslash}m{3cm}}

\usepackage[section]{placeins}
\usepackage{subcaption}
\usepackage{longtable} 
\usepackage[utf8]{inputenc}
\usepackage{lipsum} % just for dummy text- not needed for a longtable
\usepackage{lscape}
\usepackage{fixmath}
\usepackage{amssymb}
\usepackage{algpseudocode}
\usepackage{varwidth}
\usepackage[linesnumbered,ruled,vlined]{algorithm2e} 
\usepackage{comment}

\begin{document}

\title{A Multi-Lead Fusion Method for the Accurate Delineation of QRS Complex Location in 12 Lead ECG Signal}
\author{Chhaviraj~Chauhan,~\IEEEmembership{Research~Scholar,~Bharti~School,~IIT,~Delhi,}
      Monika~Agrawal,~\IEEEmembership{Professor,~CARE,~IIT~Delhi,~New~Delhi}
        and~

          Pooja~Shabherwal,~\IEEEmembership{Assistant~Professor,~The~NorthCap~University,~Gurgaon}
          \thanks {The manuscript is under review.
          (corresponding author email address-chhavirajchauhan@gmail.com) }
}
\onecolumn
\doublespacing
\maketitle
\begin{abstract}
This paper presents a multi-lead fusion method for the accurate and automated detection of the QRS complex location in 12 lead ECG (Electrocardiogram) signals. The proposed multi-lead fusion method accurately delineates the QRS complex by the fusion of detected QRS complexes of the individual 12 leads. The proposed algorithm consists of two major stages. Firstly, the QRS complex location of each lead is detected by the single lead QRS detection algorithm. Secondly, the multi-lead fusion algorithm combines the information of the QRS complex locations obtained in each of the 12 leads. The performance of the proposed algorithm is improved in terms of Sensitivity and Positive Predictivity by discarding the false positives. The proposed method is validated on the ECG signals with various artifacts, inter and intra subject variations.
The performance of the proposed method is validated on the long duration recorded ECG signals of St. Petersburg INCART database \cite{goldberger2000physiobank} with Sensitivity of 99.87\% and Positive Predictivity of 99.96\% and on the short duration recorded ECG signals of CSE (Common Standards for Electrocardiography) multilead database \cite{willems1987reference} with 100\% Sensitivity and 99.13\% Positive Predictivity. 
\end{abstract}

\IEEEpeerreviewmaketitle
\section{Introduction}
\IEEEPARstart{E}{arly} detection of heart diseases have become extremely important as cardiovascular disease (CVDs) are the major cause of mortality worldwide. The ECG signal is an effective non- invasive way for heart rate monitoring in a simple setting, operation room and even for the diagnosis of various heart diseases.The ECG signal is a record of electrical activity of the heart and the 12 lead ECG signal captures the electrical potential of heart from different angles. The 12 lead ECG signal shows the pathological status of cardiovascular system by changes in its waveforms or rhythms \cite{van2004clinical}. In the clinical environment, physicians analyze the ECG records of patients to judge whether they have a benign or unkind heart state \cite{chen2020crucial}. Manual analysis process is much laborious and repetitive. Hence, automated ECG signal analysis is required. In the past several decades, a plethora of research work has been published on the automatic diagnosis of the cardiovascular diseases.

For automated ECG signal analysis algorithms, the detection of QRS complexes is the first and foremost step, as it corresponds to the electrical excitation of the two ventricles. The duration, morphology and amplitude of the QRS complex provides the information about the current state of the heart. The accurate detection of the QRS complex is an essential step in the automated 12 lead ECG signal analysis, as any cardiac disorder changes the morphology of the QRS complex and the duration of the RR interval, which is considered clinically important. The precise detection of other sub waves like P wave, T wave and ST segment is dependent on the accurate detection of the QRS complex \cite{habib2019impact}.

In literature several methods have been proposed for detection of QRS complex in the ECG signal. Although most of the research groups have worked for detection of QRS complex detection in single lead\cite{pan1985real, hamilton1986quantitative, arzeno2008analysis, pal2012empirical,sabherwal2017automatic, junior2016real, jia2020robust,sharma2019accurate} or two lead\cite{chandra2018robust,liu2015dual} ECG signal. To have more precise analysis of the state of the heart, the QRS complex has to be accurately delineated in the 12 lead ECG signal, clinically. For detection of QRS complex in 12 lead ECG signals, the researchers have detected the QRS complex in single lead and then combined them using various combining methods. The popular single lead QRS detection methods used by many authors are Pan-Tompkins method \cite{laguna1994automatic,ledezma2015data,zhao2018multilead,ledezma2019optimal,ledezma2015fusion}, wavelet-based \cite{yochum2016automatic,yu2016fusion}, Hilbert transform-based \cite{ledezma2019optimal} and SVM based \cite{mehta2008combined} algorithms. For the detection of QRS complex in the 12 lead ECG signal,  combining or merging methods like voting fusion \cite{ledezma2015data,zhao2018multilead,ledezma2015fusion,yu2016fusion}, optimum data fusion \cite{ledezma2019optimal} , OR/AND \cite{ledezma2015fusion}, reliability index \cite{yochum2016automatic}, combined entropy \cite{mehta2008combined}, complex pan-tompkins wavelet \cite{thurner2021complex} and multi-lead QRS detection algorithm \cite{laguna1994automatic} are used by researchers.

In \cite{laguna1994automatic}, the QRS complex is detected on single lead by using Pan-Tomkins algorithm and then multi-lead detection algorithm is used for delineation of QRS complex on the 15 Lead ECG signal. In this work, 12 standard lead and 3 orthogonal leads XYZ were used for the evaluation. For a valid QRS detection, all 15 QRS location of each lead is required. Therefore, this method can only reduce the false QRS detection with a reduction of true QRS detection.  This algorithm was evaluated on CSE(Common Standards for Electrocardiography) multilead database \cite{willems1987reference,smivsek2017cse} having short duration ECG records, only in terms of mean and standard deviations but the sensitivity and positive predictivity is not presented. In \cite{saxena2002feature}, the QRS complex was detected on the 12 lead ECG signal but the detection on each lead was done sequentially. In \cite{mehta2008combined} Mehta et al. presented a support vector machine based 12 lead QRS detector and validated the algorithm on CSE database only. The performance of the algorithm was not evaluated on any long time recorded ECG signals as in INCART database. Saini et al. \cite{saini2013qrs} presented a KNN based method and evaluated on CSE database but the results for single lead ECG signal are only presented.

In a work done by Ledezma et al. \cite{ledezma2019optimal}, six different single lead QRS detector and optimum data fusion algorithm was used and evaluated on INCART database with good sensitivity but a supervised training period of few minutes was required for the calculation of weight coefficients associated with each lead. The work reported by Mondelo et al. \cite{mondelo2017combining} detected the QRS complex location by arranging all available QRS complex location in ascending order and then by defining two limits. The method first detects the actual QRS locations and later the false positives, which make the algorithm not suitable for real medical applications. In the work presented by Huang et al. \cite{huang2009qrs}, the Principal Component Analysis and the Combined Wavelet Entropy were used to detect QRS complex location but the method is not possible in real time application, as the full ECG signal recording is required to extract the principal components. In a recent work done by Thurner et al. \cite{thurner2021complex}, three single lead algorithms was merged and the multilead ECG signal was reduced to single ECG signal by averaging it. So it became a single lead QRS detection algorithm. However, many researchers presented various algorithms but still there are gaps to detect QRS complex accurately due to several reasons including diversity of the QRS waveform, low signal-to noise ratio (SNR), inter and intra morphological changes and artifacts accompanying ECG signals. Many of the existing algorithms are also not able to perform noise reduction and QRS complex detection simultaneously for the 12 lead ECG signal.

In this paper, an efficient multi-leads fusion algorithm is presented for the delineation of QRS complex in the 12 lead ECG signal. The QRS complex is detected in each lead of the observed 12 lead ECG signal using combination of wavelet transform, Hilbert transform and adaptive thresholding  \cite{sabherwal2017automatic}. The detected QRS complexes in all the 12 lead are fused together by the proposed multi-leads fusion method to have accurate delineation of QRS complex in the 12 lead ECG signal. The method gives average sensitivity and postive predictivity of 99.87\% and 99.96\% respectively for INCART database \cite{goldberger2000physiobank} and 100\% sensitivity and 99.13\% positive predictivity for the CSE database \cite{willems1987reference}.


\section{Problem Formulation}

%\IEEEpubidadjcol
The QRS complex detection is the most important point in analysis of the state of the heart. The QRS complex location is essential for the majority of the automatic arrhythmia detection algorithm. The classification efficiency of any cardiac abnormality detection method from ECG signal is also depends on the QRS complex detection algorithm. The QRS detector with low sensitivity and positive predictivity directly degrade the classification algorithm's performance, which led to a futile ECG analysis system. The sensitivity and positive predictivity of the automatic detectors depends on the accurate detection of QRS complex. Most of the researchers have worked on the single lead ECG signal. But as, the 12 lead ECG signal captures the electrical potential of heart from different angles, which gives more precise information of the status of the heart. Therefore, here the algorithm is proposed, which automatically detects QRS complex on the 12 lead ECG signal. In this algorithm, the QRS complex is detected on each lead and then fused together intelligibly. For detection of QRS complex any single lead detector can be used and then fused together by the proposed fusion method. Here, the single lead detector proposed in \cite{sabherwal2017automatic} is used as it gives better sensitivity and positive predictivity as compared to the other detectors proposed in literature. Before discussing fusion algorithm for 12 lead ECG signals, the single lead QRS complex detection algorithm is summarized ahead. 

\subsection{\textbf{Detection of QRS in single lead}}
Mathematically, the single lead ECG signal $z[n]$ can be represented as\cite{reddy2015introduction}.
\begin{equation} \label{sig}
z[n]=s[n]+w[n]
\end{equation}
where $s[n]$ represents the ECG signal and $w[n]$  represents the artifacts present in the ECG signal. For diagnosis of the disease in the 12 lead ECG signal, the first step is to detect QRS complex location accurately. As the recorded ECG signal is corrupted by various artifacts such as baseline wandering artifact, motion artifacts, and muscles contraction artifacts. These artifacts are to be removed from each lead for accurate delineation of QRS complex from the 12 lead ECG signal.

 The initial pre-processing i.e removing of various artifacts is done using discrete wavelet transform (DWT) with db6 mother wavelet. The DWT analyzes the signal at different resolution (hence, multi resolution) through the decomposition of the signal into several successive frequency subbands. The noise content is significant in high frequency detail subbands, while most of the spectral energy lies in low frequency subbands \cite{castillo2013noise}. As the energy spectrum of db6 mother wavelet is concentrated at low frequencies and its shape resembles the QRS complex,  therefore it has turned out to be the best choice among other wavelet functions for this problem.  \cite{polikar1996wavelet}.
 The DWT is performed on the observed ECG signal $z[n]$ by passing it through high pass filter $h[n]$ and a low-pass filter $l[n]$ with a down sampling factor of 2 as in \eqref{di} and \eqref{ai}:
 
 \begin{equation} \label{di}
d_{i}[n]=\sum_{k} z[k] \cdot h[2 n-k]
\end{equation}
  \begin{equation} \label{ai}
a_{i}[n]=\sum_{k} z[k] \cdot l[2 n-k]
\end{equation}
 where $i \in  I$, $d_{i}[n]$ and $a_{i}[n]$ are detail and approximation coefficients. The approximation and detail information
obtained till level 10 are labeled as $a_1[n]$ to $a_{10}[n]$ and $d_1[n]$ to $d_{10}[n]$, respectively. As the QRS complex of a ECG signal lies in a frequency band of 5 to 25 Hz \cite{elgendi2009r}, the details coefficients $d_4$ and $d_5$ are only considered \cite{sabherwal2017automatic} and the estimated ECG signal $s_0[n]$ is obtained as the sum of $d_4$ and $d_5$. 
To further attenuate the non-QRS region, the reconstructed signal $s_0[n]$ is passed through the following low pass filter \cite{sabherwal2017automatic}. 
\begin{equation} \label{slp}
           s_{LP} [n]=  \frac{1}{8}\sum_{n_0=1}^{7} \frac{12}{5} s_0 [n-n_0 ]-\frac{12}{5} s_0 [n-n_0-2) ]-\frac{11}{10} s_0 [n-n_0-4]
\end{equation}
After low pass filtering, the hilbert transform of the signal $s_{LP} [n]$ is performed to have the envelope of the QRS complex as in \eqref{shn},
\begin{equation} \label{shn}
s_H [n] = H(s_{LP} [n])	
\end{equation}
The QRS complex obtained in \eqref{shn} $s_H [n]$ is further accentuated as $s_{nl} [n]$ by applying non linear transform as,
\begin{equation} \label{snl}
	s_{nl} [n] =\frac{1}{\left|{s_H [n]}\right|} s_H^2 [n]  	
\end{equation}
The enhanced peaks $R_{Loc}$ are accurately delineated by applying adaptive thresholding \cite{sabherwal2017automatic},
\begin{equation}
	R_{Loc}= \mathcal{T}_a (s_{nl} [n]) 
\end{equation}
Where $R_{Loc}$ has the locations of detected QRS complex for a single lead ECG signal obtained by adaptive thresholding.

\subsection{\textbf{Detection of QRS complex in 12 Lead ECG signal}}
The delineation of QRS complex is done for all the 12 leads by the  aforementioned method and the detected QRS complexes for each lead are labeled as:
\begin{equation}
R_{Loc}{}_l=[R_l(1),R_l(2),R_l(3)...R_{l}(N_l)] 
\end{equation}
where $l$ denotes the lead index and $N_l$ is the number of detected QRS complex in $l^{th}$ lead. Hence, the detected QRS complexes for $L$ lead ECG signal can be represented as:

\begin{equation} \label{matrix}
\begin{bmatrix}
R_{Loc}{}_1\\
R_{Loc}{}_2\\
:\\
R_{Loc}{}_L 
\end{bmatrix}=
\begin{bmatrix}
R_1(1) & R_1(2) &......& R_1(N_1)\\
R_2(1) & R_2(2) &......& R_2(N_2)\\
: & : & ......& :\\
R_L(1) & R_L(2) &......& R_L(N_L)\\ 

\end{bmatrix}
\end{equation}

Here, $L$ is used as the proposed algorithm can give results for any combination of $L$ leads. We have validated our results with $L=12$. Practically, the number of detected QRS complex may be different for each lead, as the QRS detector might give some false alarm or might miss some points in a particular lead. So, to get a final accurate QRS complex, a fusion algorithm is required which merges the detected QRS complex of each lead belonging to the same cycle and ignores the false detection obtained in each lead.

\begin{figure}[hbt!]
 \includegraphics[width=0.7\textwidth,center]{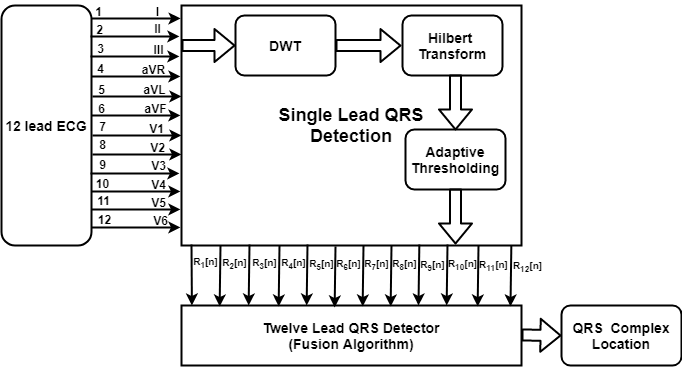}
\caption{An overview of 12 lead QRS complex detection method }
\label {overview}
\end{figure}

\subsection{\textbf{Proposed Fusion method}}
 The QRS complex locations detected in \eqref{matrix} are combined using the proposed fusion method for accurate delineation of QRS complex in the $L$ lead ECG signal. For a better estimate of QRS complex location, the proposed algorithm fuses the detected QRS complex locations corresponding to the same cardiac cycle of each lead.  The biggest challenge with this algorithm is to ascertain that the detected QRS complex belong to the same cardiac cycle. As each lead will have its own physiology, and the corresponding QRS complex of particular cardiac cycle in all leads might not be detected at exactly the same sequence. Moreover, while detecting QRS complex there might be some false negatives, or some low amplitude QRS complexes which are missed by algorithm.
 Therefore a robust algorithm is required for accurate delineation of QRS complex in $L$ lead ECG signal. The algorithm must be able to identify outliers  and missed detection, before fusion of detected beats in the $L$ leads.
Clinically, physicians also use all nearby QRS complexes to ascertain the position of the QRS complex corresponding to the given beat. To demonstrate this fact, a fusion algorithm has been suggested for accurate delineation of QRS complex, which is discussed ahead.

Let $R[n]$ = $[R_1(n), R_2(n), R_3(n),....R_L(n)]$, QRS complex vector, be the detected QRS complexes corresponding to the $L$ leads. The $[R_1(n), R_2(n), R_3(n),....R_L(n)]$ in $R[n]$ may correspond to the $n$th beat. 
%Algorithm needs to ascertain this belongingness of detected QRS complexes.  

The first step is to arrange the detected QRS complex vector $R[n]$ in the ascending order as,
\begin{equation} \label{fn}
    \textbf{f}[n] \triangleq sort(R[n])  =[f_1(n), f_2(n),f_3(n),...f_L(n)]
\end{equation}
 where \textbf{f}[$n$] indicates the sorted values of QRS complex. $QRS_{min}[n]$ and $QRS_{max}[n]$ is the minimum and maximum value of the ascended QRS complex sequence $\textbf{f}[n]$ as,
\begin{equation} \label{qmin}
QRS_{min}[n] \triangleq min({\textbf{f}[n]}) = f_1(n)  %min(QRS_1(N)) = min(R_1_1, R_2_1, R_3_1,....R_L_1)
\end{equation}
\begin{equation} \label{qmax}
QRS_{max}[n] \triangleq max({\textbf{f}[n]}) = f_L(n) %max(QRS_1(N)) = max(R_1_1, R_2_1, R_3_1,....R_L_1)
\end{equation}

 Two local arrays $\mathbold{\alpha}_1[n]$ and $\mathbold{\alpha}_2[n]$ are defined as,

\begin{equation} \label{a1}
\mathbold{\alpha}_{1}[n] = \textbf{f}[n] : f_i[n] \in [QRS_{min}[n], (QRS_{min}[n]+\delta)]\} \hspace{0.35cm} 1\leq i \leq L
\end{equation}
\begin{equation} \label{a2}
\mathbold{\alpha}_2[n] = \textbf{f}[n] : f_i[n] \in [(QRS_{max}[n]-\delta), QRS_{max}[n]]\} \hspace{0.35cm} 1\leq i \leq L
\end{equation}

where $\delta $ should lie between 80ms to 100ms which is the time span of QRS complex clinically \cite{pal2013textbook}. Therefore a mean value $\delta = 90ms $ is used in this work.

Before proceeding, we are defining few terms for clarity: 
\itemize
\item \textbf{Equivalent arrays:} The two arrays $\mathbold{\alpha}_1[n]$ and $\mathbold{\alpha}_2[n]$ are equivalent if their cardinality is equal, i.e.
\begin{equation} \label{equiv}
%\begin{align*}
\begin{split}
     %\text{cardinality of }  \mathbold{\alpha_1[n]} (u) &= \text{cardinality of } \mathbold{\alpha_2[n]}(v) \hspace{0.5cm} and,\\
     \textbf{card}(\mathbold{\alpha}_1[n]) &= u = \textbf{card}(\mathbold{\alpha}_2[n]) = v \hspace{0.5cm}               
\end{split}
%\end{align*}
\end{equation}
where $u$ and $v$ are the cardinality of $\mathbold{\alpha}_1[n]$ and $\mathbold{\alpha}_2[n]$ respectively. 
\item \textbf{Identical arrays:}
Further, if two arrays are equivalent i.e. have same cardinality and have same elements than the two arrays $\mathbold{\alpha}_1[n]$ and $\mathbold{\alpha}_2[n]$ are said to be identical. Mathematically, identical arrays are defined as:

\begin{equation} \label{ident}
%\begin{align*}
\begin{split}
     %\text{cardinality of }  \mathbold{\alpha_1[n]} (u) &= \text{cardinality of } \mathbold{\alpha_2[n]}(v) \hspace{0.5cm} and,\\
     \textbf{card}(\mathbold{\alpha}_1[n]) &= u = \textbf{card}(\mathbold{\alpha}_2[n]) = v \hspace{0.5cm} and, \\                          \mathbold{\alpha}_1^j[n]&=\mathbold{\alpha}_2^j[n] \hspace{0.5cm} \forall j, \hspace{0.5cm} 1\leq j \leq u,v
\end{split}
%\end{align*}
\end{equation}

where, j is the index for elements in $\mathbold{\alpha}_1[n]$ and $\mathbold{\alpha}_2[n]$. 
\item \textbf{Absolutely identical arrays:}
The identical arrays with cardinality $L$ are known to be absolutely identical arrays.

\item \textbf{Equivalent but not identical arrays:} That means, cardinality of both arrays are equal and their elements are not same,
\begin{equation} \label{Eqnid}
%\begin{align*}
\begin{split}
     %\text{cardinality of }  \mathbold{\alpha_1[n]} (u) &= \text{cardinality of } \mathbold{\alpha_2[n]}(v) \hspace{0.5cm} and,\\
     \textbf{card}(\mathbold{\alpha}_1[n]) &= u = \textbf{card}(\mathbold{\alpha}_2[n]) = v \hspace{0.5cm} and, \\                          \mathbold{\alpha}_1^j[n]&\neq \mathbold{\alpha}_2^j[n] \hspace{0.5cm} \forall j, \hspace{0.5cm} 1\leq j \leq u,v
\end{split}
%\end{align*}
\end{equation}

\item \textbf{Neither equivalent nor identical arrays:} That means, cardinality of both arrays are not equal and their elements are not same,
\begin{equation} \label{nEqnid}
%\begin{align*}
\begin{split}
     %\text{cardinality of }  \mathbold{\alpha_1[n]} (u) &= \text{cardinality of } \mathbold{\alpha_2[n]}(v) \hspace{0.5cm} and,\\
     \textbf{card}(\mathbold{\alpha}_1[n]) &\neq \textbf{card}(\mathbold{\alpha}_2[n])  \hspace{0.5cm} and, \\                          \mathbold{\alpha}_1^j[n]&\neq \mathbold{\alpha}_2^j[n] \hspace{0.5cm} \forall j, \hspace{0.5cm} 1\leq j \leq u,v
\end{split}
%\end{align*}
\end{equation}
\enditemize
The two arrays $(\mathbold{\alpha}_1[n]$ and $\mathbold{\alpha}_2[n])$  generated in \eqref{a1} and \eqref{a2} are compared to ascertain that QRS vector $R(n)$ belong to the same cardiac cycle. This leds to the following four cases: %as shown in flow graph of Fig \ref{flowgraph}: 

\textbf{Case 1:} \textbf{Arrays $\mathbold{\alpha}_1[n]$ and $\mathbold{\alpha}_2[n]$ are absolutely identical:} That means the cardinality of both arrays is $L$ and elements of both the arrays are equal. This
should imply that all the $L$ detected beats $(R_1[n], R_2[n], R_3[n],....R_L[n])$ are of same cardiac cycle.

\textbf{Case 2:} \textbf{Arrays $\mathbold{\alpha}_1[n]$ and $\mathbold{\alpha}_2[n]$ are identical:} The cardinality of both arrays is $K$($\leq K < L$) but less than total number of leads and also elements of both them are equal. This
should indicate that the $K$ out of $L$ detected beats belong to the same cardiac cycle.

\textbf{Case 3:} \textbf{Arrays $\mathbold{\alpha}_1[n]$ and $\mathbold{\alpha}_2[n]$ are equivalent but not identical:}  The cardinality of both arrays is $K$ ($1\leq K < L$) but elements of both them are not equal. This should imply that the $K$ detected beats do not belong to same cardiac cycle.

\textbf{Case 4:} \textbf{Arrays $\mathbold{\alpha}_1[n]$ and $\mathbold{\alpha}_2[n]$ are neither equivalent nor identical:}  The arrays have different cardinality and even the elements of the both arrays are not equal. This should indicate that all the  detected beats in QRS complex vector $(R[n])$ do not belong to the same cardiac cycle.

 These four cases are discussed ahead in detail, 

\textbf{Case 1:} When the generated arrays $\mathbold{\alpha}_1[n]$ and $\mathbold{\alpha}_2[n]$ are absolutely identical i.e. the cardinality of both the arrays is $L$ and all the entries both the arrays are equal. It deduces that all $L$ detected beats belong to the same cardiac cycle. Hence, all these beats can be combined together to ascertain the correct delineation of QRS complex. The accurate localization of detected QRS complex can be obtained by combining the corresponding beats of all the $L$ leads by some suitable metric like mean, median and mode. Among these statistical parameters, median of the detected beats as defined in \eqref{QlocL} is a better measure of central tendency as compared with mode or mean for 
skewed distributed data \cite{jim2019introduction} as the outliers  have a smaller effect on the median.

\begin{equation} \label{QlocL}
\mathrm{QRS_{loc}[n]} = 
\frac{\mathrm{R}_ \frac{{(\mathrm{L}})}{2}[n]+\mathrm{R}_ \frac{{(\mathrm{L}+2})}{2}[n]}{2}
\end{equation}

\begin{figure}[h]
\centering
 \includegraphics[width=1.1\textwidth,center]{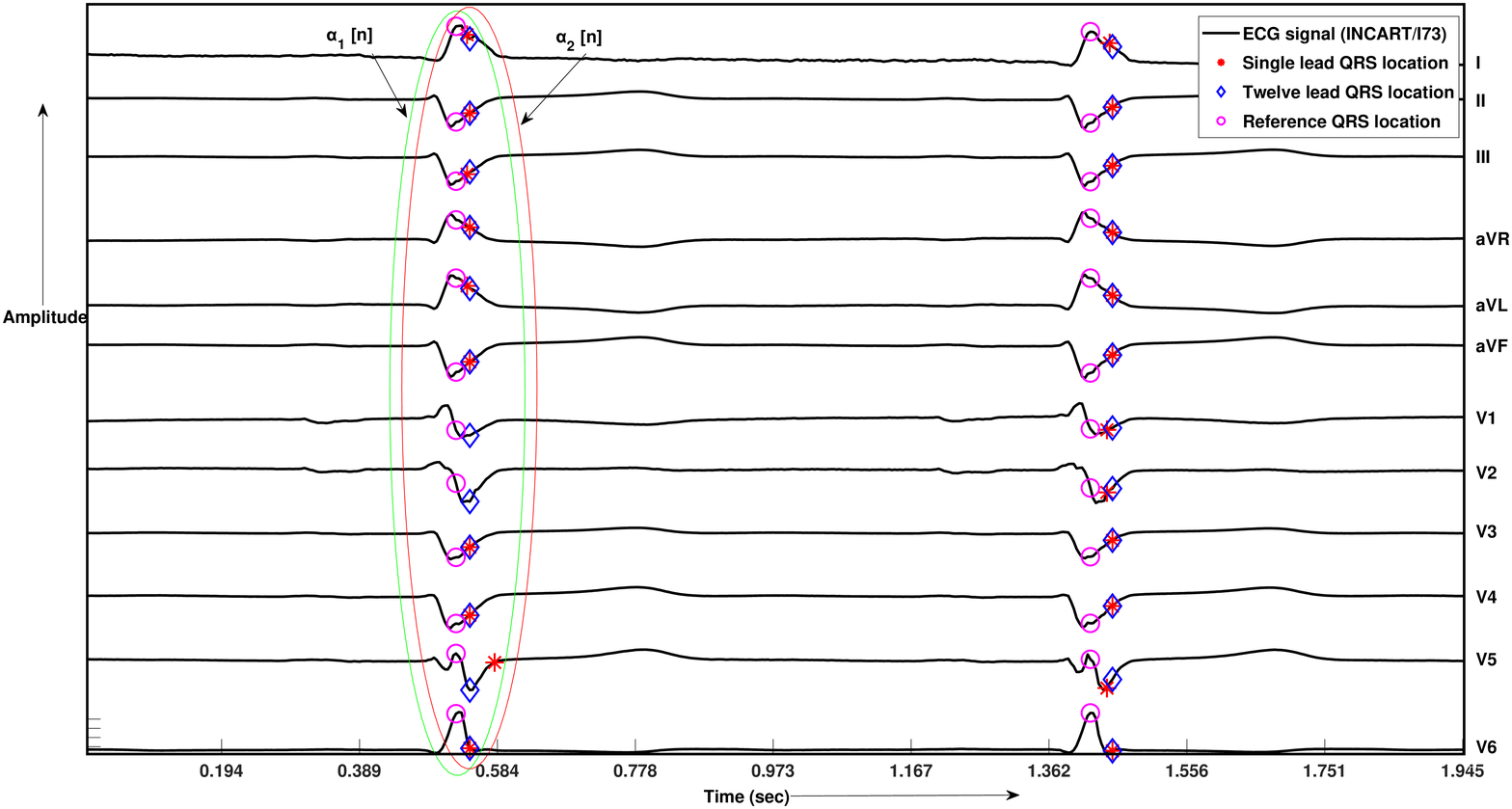}
\caption{QRS complex detection by single-lead detector and multi-lead detector in I73 data of INCART database}
\label {i73}
\end{figure}

 For better understanding, results corresponding to case 1 for the record I73 of INCART database are depicted in Fig \ref{i73}. The QRS complex locations detected by the single lead detector are marked on all the 12 leads of the record I73 from INCART database. The arrays $\mathbold{\alpha}_1[n]$ and $\mathbold{\alpha}_2[n]$ generated for the record are marked in green and red respectively. Since both the arrays are identical, the final estimated QRS complex location is marked, which is very close to the actual QRS complex location with a localization error of 19.46ms only.

\begin{figure}[hbt!]
\centering
 \includegraphics[width=0.9\textwidth,center]{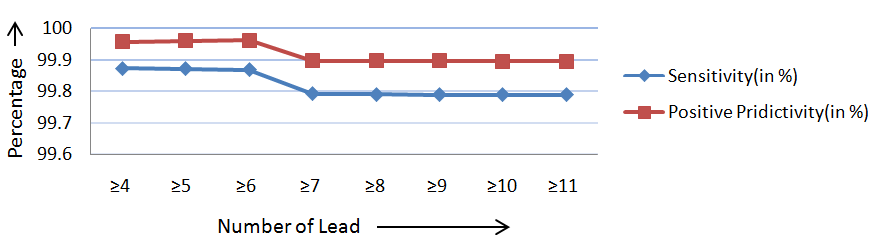}
\caption{Performance of proposed method by taking threshold of at least L number of leads}
\label {numberoflead}
\end{figure}

\textbf{Case 2:} $\mathbold{\alpha}_1[n]$ and $\mathbold{\alpha}_2[n]$ are identical i.e. the cardinality of both the arrays is $K$( $ \leq K < L$). It deduces that  $K$ detected beats belong to the same cardiac cycle. Hence, all these beats can be combined together to ascertain the correct delineation of QRS complex.   The accurate QRS complex location is estimated using \eqref{QlocN}.

\begin{equation} \label{QlocN}
 \mathrm{QRS_{loc}[n]} = \left\{\begin{array}{ll}
 \mathrm{R}_ \frac{{(\mathrm{K}+1})}{2}[n], & \text { if } \mathrm{K} \text { is odd }\\ 
 \frac{\mathbf{R}_ \frac{{(\mathrm{K}})}{2}[n]+\mathrm{R}_ \frac{{(\mathrm{K}+2})}{2}[n]}{2}, & \text { if } \mathrm{K} \text { is even }
  \end{array}\right.
\end{equation}
For this case, the minimum cardinality of both the arrays should be $K \geq 6$. As otherwise if detected beats are $<6$ no meaningful fusion is possible. Therefore to ascertain this optimum number of leads required to be combined in the proposed fusion method, an experiment for fusion is conducted for  $L$ number of leads ($4\leq$L$\leq 11$). The sensitivity and the positive predictivity was observed with varying number of leads. As shown in Fig \ref{numberoflead}, the sensitivity and positive predictivity increases with a decrease in the number of leads. This is occurring due to increased true positives with increased number of leads. There is an abrupt change in sensitivity and positive predictivity when six leads are used. Beyond this, sensitivity increases, and positive predictivity decreases slightly due to the rise in the false-positive. Hence six leads can be considered as the optimum number for the proposed fusion algorithm. %The complete algorithm is also summarized as in algorithm 3,

\textbf{Case 3:} When the cardinality of both arrays $\mathbold{\alpha}_1[n]$ and $\mathbold{\alpha}_2[n]$ is same but less than $L$ but the entries are not same i.e. temporal locations of these QRS complexes in both arrays are different. The possible reason for this case might be the minimum temporal location $QRS_{min}[n]$ and/or maximum temporal location  $QRS_{max}[n]$ are far away from the actual QRS complex location. So for this case, the common entries of both the arrays are systematically extracted. The minimum temporal location of $\mathbold{\alpha}_1[n]$ might be due to some false detection so might require replacement with the corresponding next entry and similarly the maximum temporal location of $\mathbold{\alpha}_2[n]$ might be either due to miss of the corresponding beat or detection or detection of some other wave e.g. T-wave so should be rejected. The fusion algorithm aims to make both array identical. The accurate  QRS location will be detected as in \eqref{QlocN}, once the array $\mathbold{\alpha}_1[n]$ and $\mathbold{\alpha}_2[n]$ are identical. The details of the procedure to make these arrays identical is discussed next.

\textbf{Case 4:} When $\mathbold{\alpha}_1[n]$ and $\mathbold{\alpha}_2[n]$ are not identical, i.e. either the cardinalities of arrays $\mathbold{\alpha}_1[n]$ and $\mathbold{\alpha}_2[n]$ is not same or the elements of both arrays are not equal. This infers that some false alarms have been detected and they are to discarded by the fusion algorithm. This leds to the following two sub-cases:
%The complete flow of case 2 is shown in Fig \ref{case2}
\begin{center}
\textbf{Case 4(i):} $\mathbold{card(\alpha_1[n])})$ $<$ $\mathbold{card(\alpha_2[n])}$

\textbf{Case 4(ii):} $\mathbold{card(\alpha}_1[n])$ $>$ $\mathbold{card(\alpha}_2[n])$
%\textbf{Case 2(iii):} $\mathbold{card(\alpha_1[n])}$ $=$  $\mathbold{card(\alpha_2[n])}$
\end{center}

These sub-cases are discussed ahead in detail.

\textbf{Case 4(i):} \textbf{When cardinality of $\mathbold{\alpha}_1[n]$ is less than the cardinality of $\mathbold{\alpha}_2[n]$}. This implies that the  majority of the detected QRS complex locations corresponding to all $L$ leads are close to $QRS_{max}[n]$ and therefore lie in $\mathbold{\alpha}_2[n]$. Here, one can infer that the actual QRS complex location is closer to the array $\mathbold{\alpha}_2[n]$. Also the minimum temporal location of $\mathbold{\alpha}_1[n]$ i.e $QRS_{min}[n]$ is much away from the actual QRS complex location. This may be due to the false alarm in the corresponding lead. This false detection of the concerned lead should be discarded
and replaced by the next beat of the corresponding lead to generate the correct QRS complex vector. After this replacement, a new QRS complex vector $\mathbold{R[n]}$ is generated. With this new $\mathbold{R[n]}$, again the fusion process is applied. 
The complete process is shown in the Algorithm 1. 

\begin{algorithm}[h]
\SetAlgoLined
\DontPrintSemicolon

{\textbf{Initialization:}} 
\begin{enumerate}
\item $n$ $ \rightarrow$ Index for current beat under fusion.
  \item $k$ $ \rightarrow$ Index for detected temporal locations as QRS complex in $l^{th}$ lead $(1\leq l \leq L)$.
   \item $R(n)$ $\rightarrow$ QRS complex vector for the $n^{th}$ beat.
    \item \textbf{f}[$n$] $\rightarrow$ Sorted value of $R[n]$ in ascending order.
         \item $\mathbold{\alpha}_1[n],\mathbold{\alpha}_2[n]$ $\rightarrow$ Generated arrays for $n^{th}$ beat
     \item $QRS_{min}[n]$ $\rightarrow$ Minimum temporal location in \textbf{f}[$n$] for $n^{th}$ cycle.
    \item $R_l(k)$ $\rightarrow$ $k^{th}$ temporal location belongs to lead $l$ $(k \geq n)$.
\end{enumerate}
 {\textbf{Condition:}} $card(\mathbold{\alpha_1[n]}) < card(\mathbold{\alpha_2[n]})$ 

$R_l[k] := QRS_{min}[n]$   \Comment{Find the lead of minimum temporal location in R[$n$]}

 \While{$\mathbold{\alpha_1[n]} \neq \mathbold{\alpha_2[n]}$}
 {
 $R_l(k) = R_l(k+1)$  \Comment{ Update the next temporal location in $l_{th}$ lead}\\
$\textbf{f}[$n$] \colon$ Find as in \eqref{fn} \\
$\mathbold{\alpha_1[n]}\colon$ Find as in \eqref{a1}  \\
$\mathbold{\alpha_2[n]}\colon$ Find as in \eqref{a2}\\
 }

$QRS_{loc}[n]$ =find as in \eqref{QlocN}
\caption{Algorithm for case 4 (i)}
\label{case2i}
\end{algorithm}

\begin{figure}[hbt!]
\centering
 \includegraphics[width=1.1\textwidth,center]{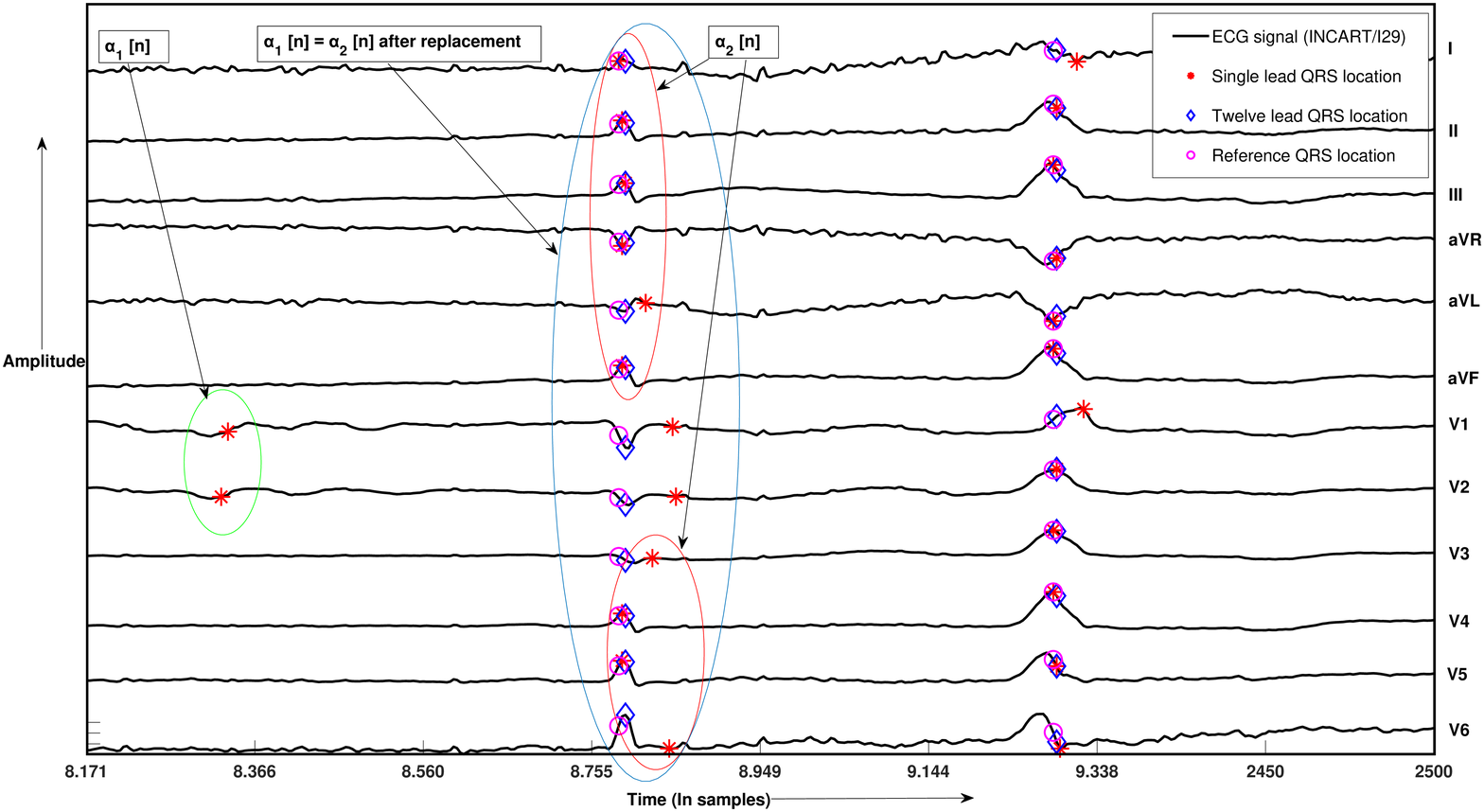}
\caption{QRS complex detection by single-lead detector and multi-lead detector in I29 data of INCART database}
\label {i29}
\end{figure}

 For the depiction of case 4(i), results corresponding to the record I29 of INCART database are shown in Fig \ref{i29}. This record corresponds to a 41 year old female having Premature Ventricular Complexes (PVCs) and Paroxysmal Ventricular Tachycardia. The QRS complex locations detected by single lead detector is marked on all the $L$ leads of the record I29.  The arrays formed by the fusion process is also marked. Since both the arrays formed are not identical as shown in Fig \ref{i29}. As $\mathbold{\alpha}_1[n]$. is less then $\mathbold{\alpha}_2[n]$, minimum temporal locations of $\mathbold{\alpha}_1[n]$ will be replaced by the next temporal locations of the same lead. The replacement of the beat led to identical arrays. The accurate QRS complex location is estimated using \eqref{QlocN}, which comes out to be at 8.794s as compared to 8.786s marked by physician with an average localization error of only 0.04s. The false positive detected at 8.335s and 8.327s locations  for lead V1 and V2 respectively are ignored by the proposed algorithm. The detected QRS complex is estimated with localization error of 8ms only, which is very close to the observed QRS complex by physician.

\begin{algorithm}[h]
\SetAlgoLined
\DontPrintSemicolon

{\textbf{Initialization:}} 
\begin{enumerate}
\item $n$ $ \rightarrow$ Index for current beat under fusion.
  \item $q$ $ \rightarrow$ Index for QRS complex location in $l^{th}$ lead $(q\geq n+1)$.
   \item $R(n)$ $\rightarrow$ QRS complex vector for the $n^{th}$ beat.
    \item \textbf{f}[$n$] $\rightarrow$ Sorted value of $R[n]$ in ascending order.
     \item $\mathbold{\alpha}_1[n],\mathbold{\alpha}_2[n]$ $\rightarrow$ Generated arrays
     \item $QRS_{max}[n]$ $\rightarrow$ Maximum temporal location in \textbf{f}[$n$].

   \item $R_l(q)$ $\rightarrow$ QRS location of lead $l$ to be considered in next beat.
  
\end{enumerate}

{\textbf{Condition:}} 
$card(\mathbold{\alpha}_1[n]) > card(\mathbold{\alpha}_2[n])$ 

$R_l[q-1] := QRS_{max}[n]$   \Comment{Find the maximum temporal location in R[n] }

 \While{$\mathbold{\alpha}_1[n] \neq \mathbold{\alpha}_2[n]$}
 {
 $R_l(q) = R_l(q-1)$  \Comment{ save the $QRS_{max}[n]$ for next cycle}\\
 $R_l[q-1] := \varnothing$\\
$\textbf{f}[n] \colon$ Find as in \eqref{fn} \\
$\mathbold{\alpha}_1[n]\colon$ Find as in \eqref{a1}  \\
$\mathbold{\alpha}_2[n]\colon$ Find as in \eqref{a2}\\
 }
 
$QRS_{loc}[n]$ =find as in \eqref{QlocN}
\caption{Algorithm for case 4(ii)}
\label{case2ii}
\end{algorithm}

\textbf{Case 4(ii):}  When the cardinality of $\mathbold{\alpha_1[n]}$ is more than the cardinality of $\mathbold{\alpha_2[n]}$. 
Here some the detected QRS complex locations corresponding to $L$ leads are much away from $QRS_{max}[n]$ whereas most of them are close to the $QRS_{min}[n]$. This implies that the actual QRS complex location is closer to the array $\mathbold{\alpha}_1[n]$ and last few entries of $\mathbold{\alpha_2[n]}$ %i.e. $QRS_{max}[n]$ other entries in close vicinity of it 
are much away from the actual QRS complex location. 
There are two possible reasons for this,  %posibilities. 

   (i) QRS complex detection has been missed for that particular lead and the entry corresponding to the QRS complex belongs to the next cycle. Which should be considered in next cycle.

   (ii) The detected beat may be the tall T wave of present cycle instead of the QRS complex.

\begin{figure}[hbt!]
\centering
 \includegraphics[width=1.1\textwidth,center]{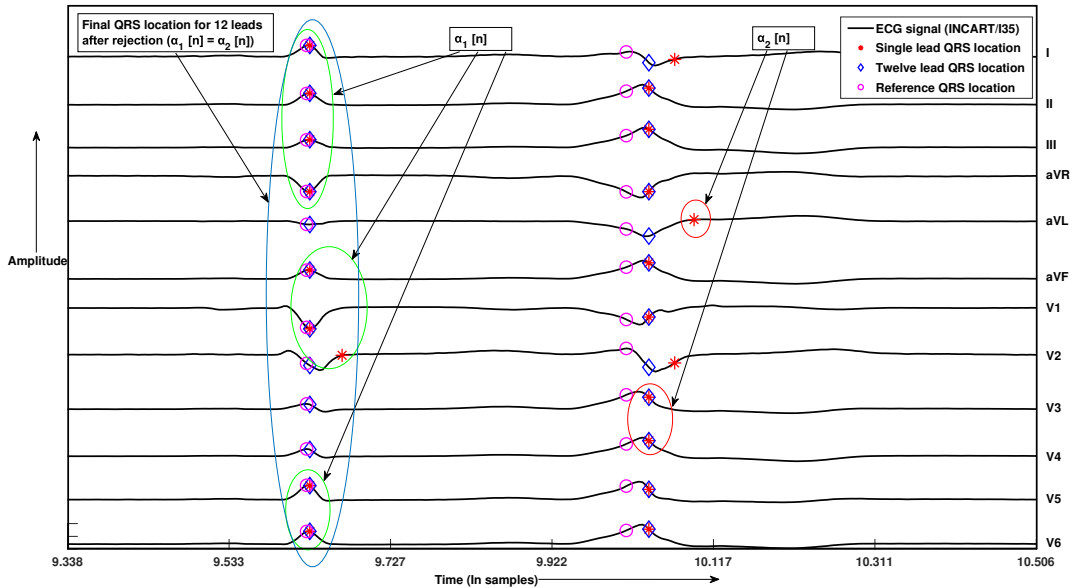}
\caption{QRS detection by single-lead detector and multi-lead detector in I35 data of INCART database}
\label {i35}
\end{figure}

In either case fusion algorithm removes corresponding false detected entries from QRS complex vector by ignoring the maximum temporal location of $\mathbold{\alpha}_2[n]$. A new QRS complex vector $R[n]$ of reduced length is obtained and the complete fusion process is again applied.
These ignored locations are then considered for the beat detection in the next cardiac cycle. This process for case 4 (ii) is also discussed in Algorithm 2.

For a depiction of case 4(ii), the record I35 of the INCART database is shown in Fig \ref{i35}. The two arrays generated by the algorithm are shown in Fig \ref{i35}. In this case, a single lead QRS detector was not able to detect the QRS location in lead aVL, V3, and V4 for the corresponding beat. Therefore the QRS locations of the subsequent beat for these leads is considered by the algorithm. Here the length of $\mathbold{\alpha}_2[n]$ is less than the length of $\mathbold{\alpha}_1[n]$. The false alarms are ignored until both arrays are identical. The false alarm at the locations 10.093s, 10.039s, and 10.039s for lead aVL, V3, and V4 are ignored. Hence, the accurate QRS complex location is estimated at 9.630s with a localization error of 3.89ms only, which is very close to the observed QRS complex by a physician. %The Sensitivity and Positive Predictivity for the record I35 come out to be 99.78\% and 100\%, respectively.
 
\begin{algorithm}[h]
\SetAlgoLined
\DontPrintSemicolon
\KwIn{$ R[n]$,$R_{loc}{}_l $ \Comment{( Detected QRS location of $n^{th}$ beat in$L$ leads and total QRS temporal locations  of $L$ leads) }}    
\KwOut{$\mathbold{\alpha}_1[n],\mathbold{\alpha}_2[n], QRS_{loc}[n] $ \Comment{(both arrays and final detected QRS location for 12 lead ECG)}}

    \SetKwFunction{FMain}{\textbf{Merge}}
    \SetKwProg{Fn}{Function}{:}{}
    \Fn{\FMain{$F$}}{
         \textbf{f}[n] = $sort(R[n])$ ;

         $QRS_{min}[n]$  $\longleftarrow$ First temporal location in \textbf{f}[n] ;

         $QRS_{max}[n]$  $\longleftarrow$ Last temporal location in \textbf{f}[n];

        $\mathbold{\alpha}_1[n]$ = Temporal location within $\delta$ limit of $QRS_{min}[n]$

        $\mathbold{\alpha}_1[n]$ = Temporal location within $\delta$ limit of  $QRS_{max}[n]$
         
        \While{$\mathbold{\alpha}_1[n] \neq \mathbold{\alpha}_2[n]$}{
        %{\eIf{$\mathbold{\alpha_1[n]} < \mathbold{\alpha_2[n]}$}
           \uIf{$card(\mathbold{\alpha}_1[n]) < card(\mathbold{\alpha}_2[n])$}
           {Replace $QRS_{min}[n]$ by next detected temporal location in same lead}
           \uElseIf{$card(\mathbold{\alpha}_1[n]) > card(\mathbold{\alpha}_2[n])$}
           {Ignore  $QRS_{max}[n]$ for next cycle }
      \Else
            {Replace $QRS_{min}[n]$ and Ignore  $QRS_{max}[n]$ 
            }
        }
            $QRS_{loc}[n]$ $=$ median of $\mathbold{\alpha}_1[n]$ or $\mathbold{\alpha}_2[n]$
        }
   \textbf{End Function}
\caption{Fusion algorithm for the QRS complex detection in 12 lead ECG signals}
\label{NagetionAlgo}
\end{algorithm}

The complete fusion algorithm for the detection of QRS complex in 12 lead ECG signal is summarized as in Algorithm 3.

%%%%%%%%%%%%%%%%%%%%%%%%%%%%%%%%%%%%%%%%%%%%%%%%%%%%%%%%%%%%%%%%%%%%%%%%%%%%%%%%%%%%%%%%%%%%%%%%%%
%Result and discussion
%%%%%%%%%%%%%%%%%%%%%%%%%%%%%%%%%%%%%%%%%%%%%%%%%%%%%%%%%%%%%%%%%%%%%%%%%%%%%%%%%%%%%%%%%%%%%%%%%%

\section{Database and Experimental Results}
\subsection{\textbf{ECG databases:}}
The proposed algorithm is evaluated on the St. Petersburg Institute of Cardiological Technics 12-lead Arrhythmia Database (INCART Database)\cite{goldberger2000physiobank} and the Common Standards for Electrocardiography (CSE) multilead database \cite{willems1987reference}. The INCART database consist of 75 annotated twelve lead ECG recording of length 30 min, each sampled at 257 Hz. The total number of beat annotation in INCART database are 175,000. The CSE set 3 database consist of 125 records of length 10 second sampled at 500Hz. The CSE multi-lead database have 15 lead simultaneously recorded ECG signals, in which 12 standard leads (I, II, III, aVR, aVL, aVF, V1, V2, V3,V4, V5 and V6) and 3 orthogonal lead(X,Y and Z). Both database have a variety of waveforms with artifacts that an arrhythmia detector might encounter in routine clinical use like sinus node function, coronary artery disease and accute myocardial infarction (MI). %The performance of the proposed algorithm is assessed using Sensitivity (Se) and Positive Predictivity (+ Pr).

\subsection{\textbf{Performance evaluation:}}
The performance of the proposed algorithm is assessed using the following statistical measures as,

\textbf{Sensitivity:} Percentage of true beats that are correctly detected by the algorithm.
 \begin{equation} \label{Sensitivityy}
Se=\frac{TP}{TP+FN}
\end{equation}

\textbf{Positive predictivity:} Percentage of detected beats that is true beats
\begin{equation} \label{Predictivity}
+Pr=\frac{TP}{TP+FP}
\end{equation}

Where \textbf{TP (true positives):} Actual QRS complex locations detected as QRS Complex,

      \textbf{FN(false negatives):} Actual QRS complex locations which have not been detected, and

      \textbf{FP(false positives):} non-QRS complex locations detected as QRS complex. 

\begin{figure} [h]
     \centering
     \begin{subfigure}[b]{0.40\textwidth}
         \centering
         \includegraphics[width=\textwidth]{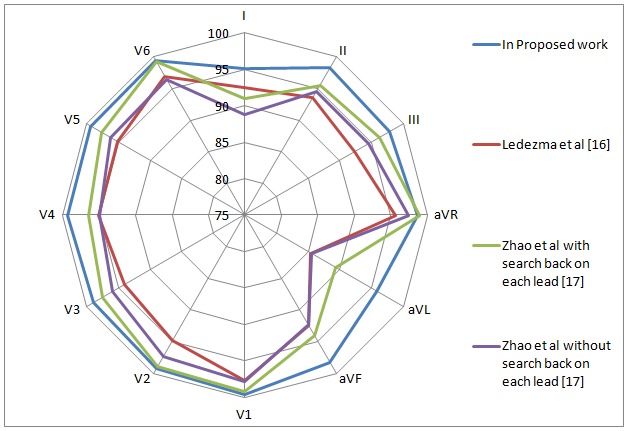}
         \caption{Sensitivity}
         \label{Sens}
     \end{subfigure}
     \hfill
     \begin{subfigure}[b]{0.41\textwidth}
         \centering
         \includegraphics[width=\textwidth]{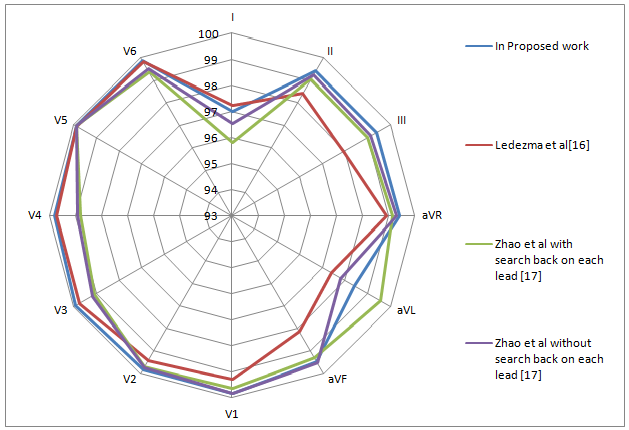}
         \caption{Positive predictivity}
         \label{Pos_pred}
     \end{subfigure}
        \caption{Performance of single lead QRS detector on 12 lead INCART database}
        \label{singlelead}
\end{figure}

\begin{figure}
     \centering
     \begin{subfigure}[b]{.8\textwidth}
         \centering
         \includegraphics[width=\textwidth]{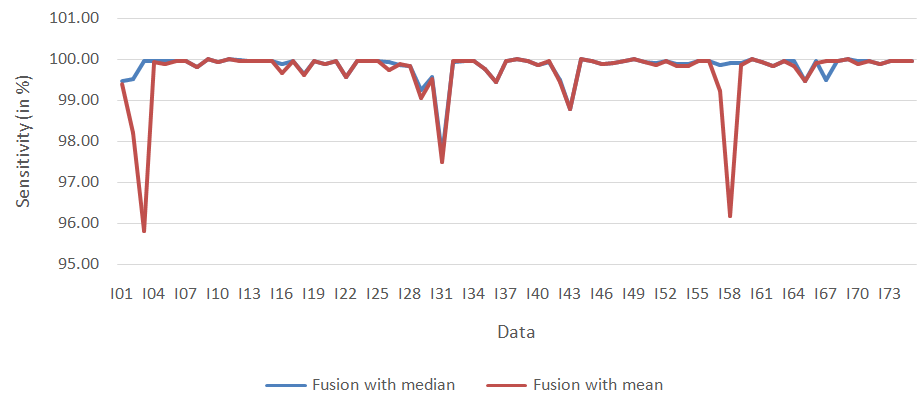}
         \caption{Sensitivity}
         \label{Se_comp}
     \end{subfigure}
     \hfill
     \begin{subfigure}[b]{0.8\textwidth}
         \centering
         \includegraphics[width=\textwidth]{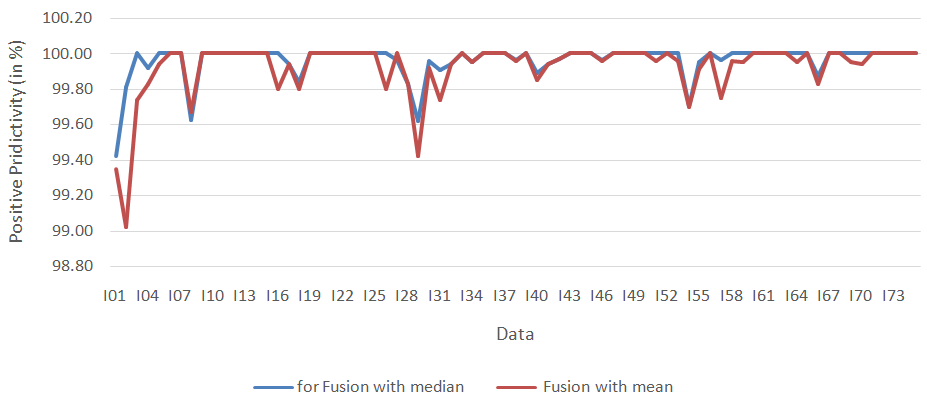}
         \caption{Positive Pridictivity}
         \label{Pr_Pr}
     \end{subfigure}
        \caption{Comparison of median-fusion and mean-fusion method on INCART database}
        \label{Mean_median}
\end{figure}

 The algorithm's performance is validated through the two standard twelve lead ECG databases, namely St. Petersburg INCART database \cite{goldberger2000physiobank} and CSE database \cite{willems1987reference}.
Initially, the single lead QRS complex detection algorithm is evaluated on 12 leads of the INCART database. The comparative results of average sensitivity and positive predictivity with the reported results in literature \cite{ledezma2015data, zhao2018multilead} are shown in Fig \ref{singlelead}. It can be seen from Fig \ref{singlelead} that the performance of the proposed algorithm presented in this paper is much better as compared to the other methods reported in the literature. The method gives the average sensitivity (Se\%) of 99.55\% and average positive predictivity (+Pr\%) of 99.87\% in lead V1, which is higher than the methods reported in the literature for the same number of leads. After obtaining the QRS complex in all $L$ leads independently by a single lead QRS complex detector, the proposed fusion method is applied. As discussed in the fusion method, the median of the all valid QRS complex location, satisfying the fusion condition, is taken instead of the mean to merged the $L$ lead. Experimentally, we observed better results when the median was taken of the $L$ detected beats compared to the mean and mode. The performance of the fusion method on 75 records of the INCART database in terms of sensitivity and positive predictivity for median and mean is shown in Fig \ref{Mean_median}. It can be seen that results obtained for the median are much better. Even the physicians annotate the QRS complex in the middle of QRS onset, and endpoint \cite{yochum2016automatic}.

\begin{figure}[hbt!]
\centering
 \includegraphics[width=1.1\textwidth,center]{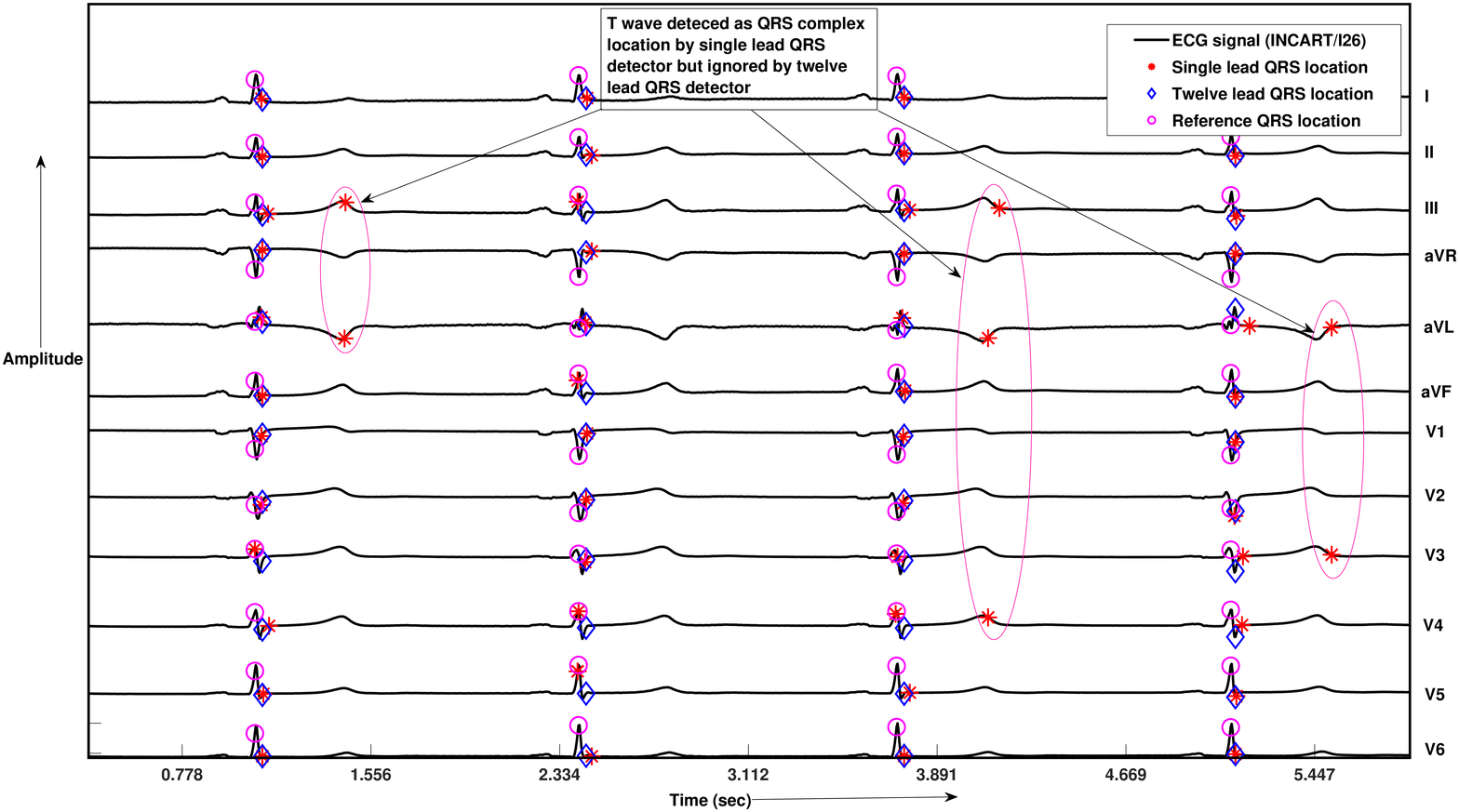}
\caption{QRS detection in record I26 of INCART database}
\label {i26}
\end{figure}

The research work on QRS detection available in literature allows a window up to 320 ms, which exceeds the maximum allowable tolerance (150 ms) for QRS detection accuracy allowed by the ANSI/AAMI-EC38, EC57 standards.
The EC38 and EC 57 standards are strictly followed in this work while reporting the results, as the bxb program of the WFDB toolbox is used. The bxb is an ANSI/AAMI-standard beat-by-beat annotation comparator \cite{silva2014open}, which matched up the detected QRS complex location with reference QRS complex annotations within the tolerance limit. The proposed method gives better performance even after including data I02, I03, I57, and I58 compared to other methods with sensitivity (Se\%) of 99.87\% and positive predictivity (+Pr\%) of 99.96\% as shown in Table III. Hence, the proposed method works satisfactorily in the absence or low signal in any lead due to bad contact of the electrode.  
 As shown in Fig \ref{i26}, the tall T waves were detected as QRS complex by a single lead QRS detector, but it was ignored by the proposed twelve lead QRS complex detector. The $L$ lead QRS detector detects the QRS complex locations with 100\% sensitivity and positive predictivity by discarding the FP detected by a single lead QRS detector. As 100\% sensitivity and positive predictivity of QRS complex detector are required for accurate MI detection, our proposed method gives the desirable results on most of the ECG signals. In Table III, the record I56 of the INCART database corresponds to a 74-year-old male with have earlier MI. The proposed method detected all the QRS locations with 100\% sensitivity and positive predictivity in this case.

 %\begin{figure}[hbt!]
\begin{figure}[hbt!]
\centering
 \includegraphics[width=1.1\textwidth,center]{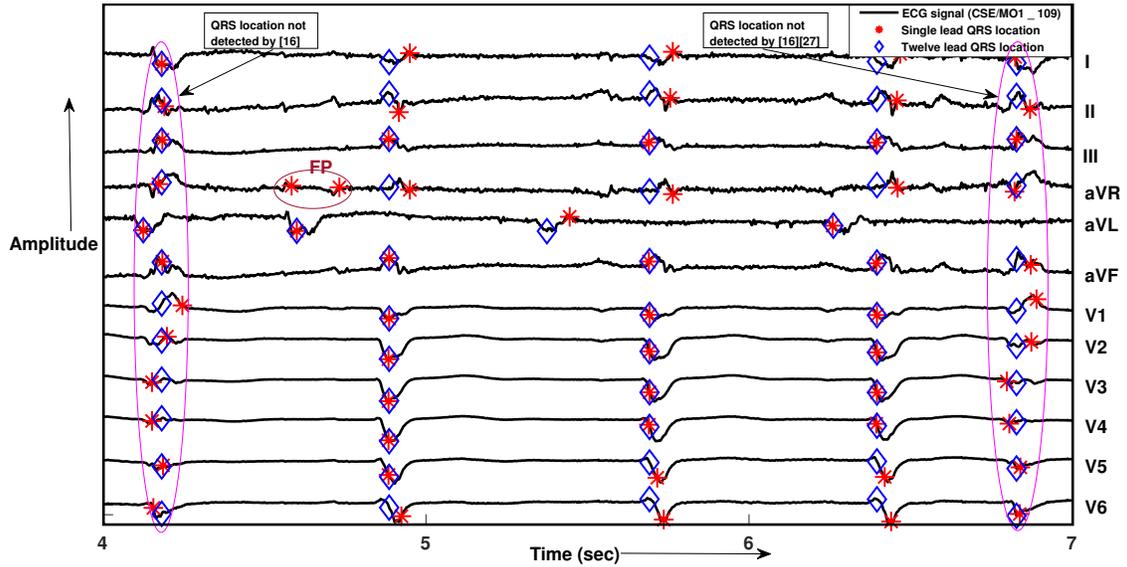}
\caption{QRS detection in record $MO1\_109$ of CSE database}
\label {cse}
\end{figure}

The $L$ lead QRS detector is also evaluated on the CSE database. The proposed method gives the sensitivity of 100\% and positive predictivity 99.13\% on the CSE database, which is better than other methods. 
%The results obtained for signals of the CSE database are shown in Table II.  
It can be seen that the fusion of all leads improves the QRS complex location.  In Fig \ref{cse}, the signal M0\_109 of the CSE database is shown, in which the QRS complex locations not detected by \cite{laguna1994automatic, saini2013qrs} are detected by the proposed algorithm. The comparative results of average Sensitivity and Positive Predicitvity with the reported results in literature \cite{thurner2021complex,ledezma2019optimal,saini2013qrs,mehta2009identification, mondelo2017combining,mehta2007development, huang2009qrs, vitek2009wavelet} are shown in Table \ref{Table_comp}.  Hence, the proposed method shows satisfactory performance on both the databases. The method reported in \cite{ledezma2019optimal} require a supervised training period to learn the weights of each lead. Therefore it can not be used for short duration ECG recordings of CSE database have 10 sec of duration. The method given by \cite{huang2009qrs} requires the full ECG signal to extract the principle components. Therefore, this can not be used for the real-time medical devices. In contrast, the proposed multi-lead fusion algorithm act as the performance booster of any single lead QRS complex detector. Hence, our method is suitable for the automation of a Holter device for real-time monitoring as well as for recorded ECG signal. The proposed algorithm is implemented in MATLAB on a i7 processor. The processing time to detect QRS complex location in 12 lead ECG signals of INCART database is given in Table \ref{Time}.

  % \begin{table}[hbt!] 
\begin{table}[ht]
\renewcommand{\arraystretch}{1.0}
\setlength{\arrayrulewidth}{0.1mm}
\setlength{\tabcolsep}{4pt}
\caption{Comparison of proposed 12 lead QRS complex detector with existing methods}
\label{Table_comp}
%\centering
\scalebox{1.1}{\begin{tabular}{cccccccc}
\hline

Database & Method&No. of beat& TP& FN&FP&Se  &+Pr  \\

\hline
INCART&Proposed&175906&175628&278&75&99.87&99.96 \\
&Thurner et al (2021) \cite{thurner2021complex}&N/R$^*$&175130&789&738&99.55&99.58\\
&Ledezma et al (GQRS)(2019)\cite{ledezma2019optimal} &147570&147469&101&186&99.94&99.88\\
&Ledezma et al(PT) (2019)\cite{ledezma2019optimal}& 175906& 175738& 168& 157& 99.91& 99.92\\
&Mondelo et al (2017) \cite{mondelo2017combining}&$\approx$81517&N/R$^*$&N/R$^*$&N/R$^*$&99.86&99.98\\
&huang et al (2009) \cite{huang2009qrs}&175900&175863&35&6&99.98&99.99\\
\hline
CSE&Proposed&1488&1488&00&13&100&99.13\\
&Mehta et al (2009)\cite{mehta2009identification} & 1488&1484&04&24&99.73&98.40\\
&Saini et al (2013) \cite{saini2013qrs} &1488&1486&02&02&99.86&99.86\\
&Vitek et al (2009) \cite{vitek2009wavelet}&N/R$^*$&N/R$^*$&N/R$^*$&N/R$^*$&99.19&N/R$^*$\\
&Mehta et al (2007) \cite{mehta2007development}&1488&1487&01&13&99.93&99.13\\
  
\hline
\end{tabular}
}
\footnotesize{$^*$ Not Reported}
\end{table}

%%%%%%%%%%%%%%%%%%%%%%%%%%%%%%%%%%%%%
\begin{table}[ht]
\renewcommand{\arraystretch}{0.8}
\setlength{\arrayrulewidth}{0.2mm}
\setlength{\tabcolsep}{6pt}
\caption{Processing time of the proposed method}
\label{Time}
\centering
\scalebox{0.9}{\begin{tabular}{|c|c|ccc|c|}
\hline

S.No.& Algorithms & &Processing time (in second)& & Duration of \\
&&Mean&Variance&Standard deviation& ECG recording\\

\hline

 1&Multi-Lead Fusion &0.6599&0.0933&0.3055&30 min\\
 2&QRS detector and&28.3626&0.2266&0.4761&30 min\\
& Multi-Lead Fusion&&&&\\
\hline
\end{tabular}
}
%\footnotesize{$^*$ Not Reported}
\end{table}

%\textcolor{red}{The most important merit of this method is that it does not require any training period to learn the weight of each lead as \cite{ledezma2019optimal}.  Hence it can be used for a short duration ECG recording of 10 sec in CSE database.}

%and can be used for the real-time medical device. The fusion-based $L$ lead QRS detection method is suitable for the automation of a Holter device for real-time long recorded as well as a short recorded ECG signal. 

\section{Conclusion}

In this paper, a novel multi-lead fusion algorithm is designed for accurate delineation of QRS complex in the 12 lead ECG signal. The proposed algorithm is based on the QRS complex detection of single lead ECG signal. The proposed method is robust against the noise or artifacts present in the ECG signal. The detection is not affected by the change of morphology of the ECG signal. %An accurate QRS complex delineation method for 12 lead ECG signals has been proposed here. 
The effectiveness of the proposed method was tested on standard ECG databases like CSE and INCART databases. The detection performance is measured in terms of  sensitivity and positive predictivity. The results obtained are compared with the existing detectors in terms of sensitivity and positive predictivity. %The results obtained are compared with the existing detectors in terms of sensitivity and positive predictivity. %The performance of the proposed algorithm is also better in the presence of artifacts than available single or multi-lead QRS detectors. 
The proposed algorithm is suitable for automated handheld devices to detect the real-time change in ECG signal. This method can be used for real time processing of huge ECG recorded data over a long period of time and tele-medicine application.

\bibliographystyle{IEEEtran}
\bibliography{main}
\begin{comment}

\end{comment}

\appendices
%\begin{appendices}
%\section{Results of 12 lead QRS detector on INCART Database} 

\begin{table}[hp]
\renewcommand{\arraystretch}{1.1}
\setlength{\arrayrulewidth}{0.25mm}
\setlength{\tabcolsep}{8pt}
\caption{Results of 12 lead QRS detector on INCART Database}
\label{Table_INCART1}
\centering
\begin{tabular}{ccccc|ccccc}

\hline

Records & Actual& TP& FN& FP&Records & Actual& TP& FN& FP  \\
 & QRS& & & & &QRS & & &   \\
\hline
I01&	2757&	2742&	15&	16&	I39	&1775&	1775&	0&	0\\
I02&	2674&	2662&	12&	5& 	I40&	2666&	2663&	3&	3\\
I03&	2451&	2451&	0&	0&	I41&	1630&	1630&	0&	1\\
I04&	2424&	2424&	0&	2& 	I42	&3109&	3095&	14&	1\\
I05&	1776&	1776&	0&	0& 	I43	&2209&	2183&	26&	0\\
I06&	2493&	2493&	0	&0&	I44&	2495&	2495&	0&	0\\
I07&	2706&	2706&	0&	0&	I45	&1928&	1928&	0&	0\\
I08&	2131&	2128&	3&	8&  	I46&	2658&	2656&	2&	1\\
I09&	2997&	2997&	0&	0& 	I47&	1953&	1952&	1&	0\\
I10&	3682&	3680&	2&	0& 	I48&	2357&	2357&	0&	0\\
I11&	2106&	2106&	0&	0& 	I49	&2147&	2147&	0&	0\\
I12&	2809&	2809&	0	&0&	I50	&2998&	2997&	1&	0\\
I13&	2023&	2023&	0&	0& 	I51	&2777&	2776&	1&	0\\
I14	&1866&	1866&	0&	0& 	I52	&1747&	1747&	0&	0\\
I15	&2635&	2635&	0&	0& 	I53&	2262&	2261&	1&	0\\
I16	&1522&	1521&	1&	0& 	I54&	2363&	2362&	1&	7\\
I17	&1673&	1673&	0&	1& 	I55	&2166&	2166&	0&	1\\
I18	&3084&	3075&	9&	5& 	I56	&1705&	1705&	0&	0\\
I19	&2063&	2063&	0&	0& 	I57&	2871&	2869&	2&	0\\
I20& 2652&	2650&	2&  0&  I58	&2325&	2324&	1&	0\\
I21&	2184&	2184&0&	0&	I59&	2148&	2147&	1&	0\\
I22	&3126&	3113&	13&	0& 	I60	&2475&	2475&	0&	0\\
I23	&2205&	2205&	0&	0& 	I61	&1454&	1454&	0&	0\\
I24	&2571&	2571&	0&	0& 	I62	&2269&	2266&	3&	0\\
I25&1712&	1712&	0&0&	I63&	1994&	1994&	0&	0\\
I26&	1509&	1509&	0&	0&	I64	&1913&	1913&	0&	0\\
I27&	2605&	2601&	4&	1& 	I65&	2664&	2650&	14&	0\\
I28	&1717&	1715&	2&	3& 	I66	&2340&	2340&	0&	3\\
I29&	2621&	2605&	16&	10&	I67&	2975&	2960&	15&	0\\
I30&	2462&	2451&	11&	1& 	I68&	2644&	2644&	0&	0\\
I31&	3210&	3138&	72&	3& 	I69&	2169&	2169&	0&	0\\
I32	&1619&	1619&	0&	1& 	I70&	1666&	1666&	0&	0\\
I33&	1837&	1837&	0	&0&	 I71&	1670&	1670&	0&	0\\
I34&	1965&	1965&	0&	1& 	I72	&2270&	2269&	1&	0\\
I35&	3676&	3668&	8&	0&	I73	&1992&	1992&	0&	0\\
I36	&3911&	3890&	21&	0& 	I74&	2405&	2405&	0&	0\\
I37&	2461&	2461&	0&	0& 	I75	&2103&	2103&	0&	0\\
I38&	2699&	2699&	0&1& \textbf{Total}&\textbf{175906}&\textbf{175628}&\textbf{278}&\textbf{75}\\

\hline
\end{tabular}
\end{table}

%\section{Results of 12 lead QRS detector on CSE Database (prefix MO1\_ with each record}

\begin{table}[hp]
\renewcommand{\arraystretch}{1.1}
\setlength{\arrayrulewidth}{0.35mm}
\setlength{\tabcolsep}{8pt}
\begin{landscape}
\caption{Results of 12 lead QRS detector on CSE Database (prefix MO1\_ with each record)}
\label{Table_CSE}
%\centering
%\resizebox{\textwidth}{!}{%
\scalebox{0.9}{\begin{tabular}{ccccc|ccccc|ccccc}
\hline

Records & Actual & TP& FN& FP&Records & Actual& TP& FN& FP &Records & Actual& TP& FP& FN \\
 &  QRS& & & &&  QRS& & &  & &  QRS& & &  \\
\hline
001	&11&	11&	0&	0&043&	10&	10	&0&	0 &	085&	11&	11&	0&	0	\\
002&	19&	19&	0&	0&  044&	08&	08	&0&	0&	086&	09&	09&	0&	0	\\
003&	17&	17&	0&	0&  045&	13&	13	&0&	0&	087&	09&	09&	0&	0	\\
004&	16&	16&	0&	0&  046&	12&	12	&0&	0&	088&	09&	09&	0&	0	\\
005&	17&	17&	0&	0&  047&	16&	16	&0&	0&	089&	06&	06&	1&	0\\
006&	16&	16&	0&	0&  048&	10&	10	&0&	0&	090&	08&	08&	0&	0\\
007&	17&	17&	0&	0&  049&	11&	11	&0&	0&	091&	09&	09&	0&	0\\
008&	10&	10&	0&	0&  050&	08&	08	&0&	0&	092&	11&	11&	0&	0\\
009&	12&	12&	0&	0&  051&	20&	20	&0&	0&	093&	09&	09&	0&	0\\
010&	07&	07	&4&	0&  052&	15&	15	&0&	0&	094&	10&	10&	0&	0\\
011&   15&15&0&0&      053&	17&	17	&0&	0&	095&	08&	08&	0&	0\\
012&	13&	13	&1&	0&  054&	07&	07	&4&	0&	096&	08&	08&	0&	0\\
013&	12&	12	&0&	0&  055&	09&	09	&0&	0&	097&	09&	09&	1&	0\\
014&	08&	08	&0&	0&  056&	10&	10	&0&	0&	098&	11&	11&	0&	0\\
015&	06&	06	&0&	0&  057&	10&	10	&0&	0&	099&	10&	10&	0&	0\\
016&	16&	16	&0&	0&  058&	15&	15	&0&	0&	100&	15&	15&	0&	0	\\
017&	10&	10	&0&	0&  059&	08&	08	&0&	0&	101&	16&	16&	0&	0\\
018&	15&	15	&0&	0&  060&	12&	12	&0&	0&	102&	16&	16&	0&	0	\\
019&	13&	13	&0&	0&  061&	13&	13	&0&	0&	103&	11&	11&	0&	0\\
020&	22&	22	&0&	0&  062&	11&	11	&0&	0&	104&	08&	08&	0&	0	\\
021&	07&	07	&0&	0&  063&	09&	09	&0&	0&	105&	14&	14&	0&	0	\\
022&	12&	12	&0&	0&  064&	11&	11&	0&	0&	106&	10&	10&	0&	0\\\
023&	08&	08	&0&	0&  065&	12&	12&	0&	0&	107&	14&	14&	0&	0\\\
024&	09&	09	&0&	0&  066&	10&	10&	0&	0&	108&	16&	16&	0&	0\\
025&	10&	10	&0&	0&  067&	12&	12&	0&	0&109&	15&	15&	0&	0\\
026&	13&	13	&0&	0&  068&	16&	16&	0&	0&110&	15&	15&	0&	0\\
027&	14&	14	&0&	0& 	069&	13&	13&	0&	0&	111&	20&	20&	0&	0\\
028&	10&	10	&0&	0&  070&	12&	12&	0&	0&	112&	13&	13&	0&	0\\
029&	10&	10	&0&	0&  071&	14&	14&	0&	0&	113&	17&	17&	0&	0\\
030&	12&	12	&0&	0&  072&	11&	11&	0&	0&	114&	11&	11&	0&	0\\
031&	13&	13	&0&	0&  073&	13&	13&	0&	0&115&	20&	20&	0&	0\\
032&	14&	14	&0&	0&  074&10&10&0&0&	116&	13&	13&	0&	0\\
033&	09&	09	&0&	0&  075&	13&	13&	0&0&117&	12&	12&	0&	0\\
034&	12&	12	&0&	0&  076&	13&	13&	0&0&118&	11&	11&	0&	0\\
035&	11&	11	&0&	0&  077&	12&	12&	0&0&	119&	18&	18&	0&	0\\
036&	12&	12	&0&	0&  078&	07&	07&	0&0&	120&	09&	09&	0&	0\\
037&	13&	13	&0&	0&  079&	09&	09&	1&0&	121&	10&	10&	0&	0\\
038&	11&	11	&0&	0&  080&	09&	09&	0&0&	122&	15&	15&	0&	0\\
039&	09&	09	&0&	0&  081&	12&	13&	0&0&	123&	13&	13&	0&	0\\
040&	12&	12	&0&	0&  082&	09&	09&	1&0&	124&	11&	11&	0&	0\\
041&	11&	11	&0&	0&  083&	15&	15&	0&0&	125&	12&	12&	0&	0\\
042&	11&	11	&0&	0&  084&	10&	10&	0&0& \textbf{Total}&\textbf{1488}&\textbf{1488}&\textbf{13}& \textbf{0}\\

\hline
\end{tabular}
}
\end{landscape}
\end{table}

\end{document}